\documentclass[runningheads]{svmult}
\usepackage{epsfig}

\begin{document}
\title*{35 Years of Testing Relativistic Gravity: \protect\newline Where do we go from here?  }
\toctitle{35 Years of Testing Relativistic Gravity: 
\protect\newline Where do we go from here? }
%
%
\titlerunning{Testing Relativistic Gravity in Space}
%
\author{Slava G. Turyshev\inst{1}
\and James G. Williams\inst{1}
\and Kenneth Nordtvedt, Jr.\inst{2}
\and \\Michael Shao \inst{1}
\and Thomas W. Murphy, Jr. \inst{3}
}
\authorrunning{Slava G. Turyshev et al.}
%
%
\institute{Jet Propulsion Laboratory, 
4800 Oak Grove Drive, Pasadena, CA 91109 USA
\and Northwest Analysis, 118 Sourdough Ridge Road, Bozeman, MT 59715 USA
\and
Physics Department, University of California, San Diego,   CASS-0424, \\ 9500 Gilman Dr., La Jolla, CA 92093, USA}

\maketitle              

\begin{abstract}
This paper addresses the motivation, technology and recent results in the tests of the general theory of relativity (GR)  in the solar system. We  specifically discuss Lunar Laser Ranging (LLR), the only   technique available to test the Strong Equivalence Principle (SEP) and presently the most accurate method to test for the constancy of the gravitational constant {\it G}. After almost 35 years since beginning of the experiment, LLR is poised to take a dramatic step forward by proceeding from cm to mm range accuracies enabled by the new Apache Point Observatory Lunar Laser-ranging Operation (APOLLO) currently under development in New Mexico. This facility will enable tests of the Weak and Strong Equivalence Principles with a sensitivity approaching 10$^{-14}$, translating to a test of the SEP violation parameter, $\eta$, to a precision of $\sim 3\times 10^{-5}$. In addition,  the $v^2/c^2$ general relativistic effects would be tested to better than 0.1\%, and measurements of the relative change in the gravitational constant, $\dot{G}/G$, would be $\sim0.1$\% the inverse age of the universe.  

This paper also discusses a new fundamental physics experiment that will test relativistic gravity with an accuracy better than the effects of the second order in the gravitational field strength, $\propto G^2$. The Laser Astrometric Test Of Relativity (LATOR) will not only improve the value of the parameterized post-Newtonian (PPN) $\gamma$ to unprecedented levels of accuracy of 1 part in 10$^{8}$, it will also be able to measure effects of the next post-Newtonian order ($c^{-4}$) of light deflection resulting from gravity's intrinsic non-linearity, as well as measure a variety of other relativistic effects.  LATOR will lead to very robust advances in the tests of fundamental physics: this mission could discover a violation or extension of general relativity, or reveal the presence of an additional long range interaction in the physical law.  There are no analogs to the LATOR experiment; it is unique and is a natural culmination of solar system gravity experiments.
\end{abstract}

\section{Introduction}

Einstein's general theory of relativity (GR) began with its empirical success in 1915 by explaining the anomalous perihelion precession of Mercury's orbit, using no adjustable theoretical parameters.  Shortly thereafter, Eddington's 1919 observations of star lines-of-sight during a solar eclipse confirmed the doubling of the deflection angles predicted by GR as compared to Newtonian and Equivalence Principle arguments.  Following these beginnings, the general theory of relativity has been verified at ever-higher accuracy. Thus, microwave ranging to the Viking Lander on Mars yielded an accuracy of $\sim$0.1\% in the tests of GR \cite{[47],[54]}.  The astrometric observations of quasars on the solar background performed with Very-Long Baseline Interferometry (VLBI) improved the accuracy of the tests of gravity to $\sim$ 0.03\% \cite{[48]}. Lunar Laser Ranging (LLR),  the continuing legacy of the Apollo program, has provided $\sim$ 0.01\% verification of the general relativity via precision measurements of the lunar orbit \cite{[62],[15llr],[39],[63],[2]}. Finally, the recent experiments with the Cassini spacecraft have improved the accuracy of the tests to $\sim$ 0.003\% \cite{cassini,cassini_and,cassini_ber}. As a result, by now not only is the `non-relativistic', Newtonian regime well understood, but the first `post-Newtonian' approximation is well-studied, making general relativity the standard theory of gravity where astrometry and spacecraft navigation are concerned. 

The continued inability to merge gravity with quantum mechanics, and recent observations in cosmology indicate that the pure tensor gravity of general relativity needs modification or augmentation.  Recent work in scalar-tensor extensions of gravity that are consistent with present cosmological models \cite{[12a],[10],[15],[12n]} motivate new searches for very small deviations of relativistic gravity in the solar system at levels of 10$^{-5}$ to 10$^{-7}$ of the post-Newtonian effects or essentially to achieve accuracy that enables measurement of the effects of the 2nd order in the gravitational field strength ($\propto G^2$).  This will require a several order-of-magnitude improvement in experimental precision from present tests. At the same time, it is well understood that the ability to measure the second order light deflection term  would enable one to demonstrate even higher accuracy in measuring the first order effect, which is of the utmost importance for the gravitational theory and is the challenge for the 21st century fundamental physics. 

Because of its importance to the tests of gravitational theory, especially to the tests of the Equivalence Principle and search for possible variation of the gravitational constant, we will concentrate on the improvements to these tests expected from LLR in the very near future. We will also discuss the recently proposed LATOR mission \cite{lator} that offers a very attractive opportunity to improve fundamental tests of gravitational theory by at least 3 orders-of-magnitude.  

LLR is the only technique currently available that allows one to test for a possible Strong Equivalence Principle (SEP) violation as well as providing the best limit on the possible variation of the gravitational constant, {\it G}. In the next few months LLR is poised to take a dramatic step forward, enabled both by detector technology and access to a large-aperture astronomical telescope. The Apache Point Observatory Lunar Laser-ranging Operation (APOLLO) is a unique instrument developed specifically to improve accuracies of LLR ranges to retroreflectors on the Moon.  The project will exploit a large (3.5~m), high-quality modern astronomical telescope at an excellent site to push LLR into a new regime of multiple return photons per pulse, enabling a determination of the shape of the lunar orbit to a precision of one millimeter \cite{[35],[35a]}. As a result, APOLLO will permit improved solutions for parameters describing the Equivalence Principle, relativity theories, and other aspects of gravitation and solar system dynamics. In particular, the Equivalence Principle test would have a sensitivity approaching 10$^{-14}$, corresponding to a sensitivity for the SEP violation parameter $\eta$ of $\sim 3\times 10^{-5}$; $v^2/c^2$ general relativistic effects would be tested to better than 0.1\%; and measurements of the relative change in the gravitational constant, $\dot{G}/G$, would be $\sim0.1$\% the inverse age of the universe.

The LATOR test will be performed in the solar gravity field using optical interferometry between two micro-spacecraft  \cite{lator}.  Precise measurements of the angular position of the spacecraft will be made using a fiber coupled multi-chanelled optical interferometer on the International Space Station (ISS) with a 100 m baseline. The primary objective of the LATOR Mission will be to measure the gravitational deflection of light by the solar gravity to an accuracy of 0.1 picoradians, which corresponds to $\sim$10 picometers on a 100 m interferometric baseline. In conjunction with laser ranging between the spacecraft and the ISS, LATOR will allow measurements of the gravitational deflection by a factor of 3,000 better than is currently known. In particular, this mission will not only measure the key parameterized post-Newtonian (PPN) $\gamma$ to unprecedented levels of accuracy of one part in 10$^8$, it will also measure for the first time the next post-Newtonian order ($c^{-4}$) of light deflection resulting from gravity's intrinsic non-linearity as well as measure a number of other relativistic effects. 

LATOR will lead to very robust advances in the tests of fundamental physics:  this mission could discover a violation or extension of general relativity, or reveal the presence of an additional long range interaction in the physical law.  By testing grevity to several orders-of-magnitude higher precision, finding a violation of general relativity or discovering a new long range interaction could be one of this era's primary steps forward in fundamental physics. There are no analogs to the LATOR experiment; it is unique and a natural culmination of solar system gravity experiments.

This paper summarizes the science motivation for the precision tests of gravity and focuses on the current and near future techniques and  methods that are used to conduct gravity experiments in the solar system. It specifically outlines the methods used in the LLR tests of $\dot G$, SEP and other PPN parameters and discusses the order-of-magnitude improvement in these tests that the next-generation of LLR technique enables.  The paper also provides an overview for the LATOR experiment including a preliminary mission design. 

\section{Scientific Motivation} 
\subsection{ PPN Parameters and Their Current Limits} 

Generalizing on a phenomenological parameterization of the gravitational metric tensor field, which Eddington originally developed for a special case, a method called the parameterized post-Newtonian (PPN) metric has been developed (see \cite{[39],[38],[60],[59]}).  This method  represents the gravity tensor's potentials for slowly moving bodies and weak interbody gravity, and is valid for a broad class of metric theories including general relativity as a unique case.  The several parameters in the PPN metric expansion vary from theory to theory, and they are individually associated with various symmetries and invariance properties of the underlying theory.  Gravity experiments can be analyzed in terms of the PPN metric, and an ensemble of experiments will determine the unique value for these parameters, and hence the metric field itself.

The PPN expansion serves as a useful framework to test relativistic gravitation in the context of the LATOR mission. In the special case, when only two PPN parameters ($\gamma$, $\beta$) are considered, these parameters have clear physical meaning. Parameter $\gamma$  represents the measure of the curvature of the space-time created by a unit rest mass; parameter  $\beta$ is a measure of the non-linearity of the law of superposition of the gravitational fields in the theory of gravity. GR, which corresponds to  $\gamma = \beta$  = 1, is thus embedded in a two-dimensional space of theories. The Brans-Dicke theory is the best known of the alternative theories of gravity.  It contains, besides the metric tensor, a scalar field and an arbitrary coupling constant $\omega$, which yields the two PPN parameter values $\gamma = (1+ \omega)/(2+ \omega)$, and $\beta$  = 1.  More general scalar tensor theories yield values of $\beta$ different from one \cite{[12a]}.

PPN formalism proves to be a versatile method to plan gravitational experiments in the solar system and to analyze the data which is obtained \cite{[48],[39],[60],[59],[44],[58],[9]}.  Different experiments test different combinations of these parameters (for more details, see \cite{[59]}). The secular trend of Mercury's perihelion, when described in the PPN formalism, depends on another linear combination of the PPN parameters $\gamma$  and  $\beta$ and the quadrupole coefficient $J_{2\odot}$ of the solar gravity field:  $\lambda_\odot = (2 + 2\gamma -\beta)/3 + 0.296\times J_{2\odot}\times 10^4$. The combination of parameters $\lambda_\odot = 0.9996\pm 0.0006$, was obtained with the Mercury ranging data \cite{[31]}. The PPN formalism has also provided a useful framework for testing the violation of the SEP for gravitationally bound bodies.  In that formalism, the ratio of passive gravitational mass $M_G$ to inertial mass $M_I$ of the same body is given by $M_G/M_I = 1 +\eta U/(M_0c^2)$, where $M_0$ is the rest mass of this body and $U$ is the gravitational self-energy. The SEP violation is quantified by the parameter $\eta$, which is expressed in terms of the basic set of PPN parameters by the relation $\eta = 4\beta-\gamma-3$. Analysis of planetary ranging data recently yielded an independent determination of parameter $\gamma$ \cite{[63],[2]}: $|\gamma -1| = 0.0015 \pm 0.0021$; it also gave $\beta$  with accuracy at the level of $|\beta -1| = -0.0010 \pm 0.0012$. With LLR finding that Earth and Moon fall toward the Sun at rates equal to 1.5 parts in 10$^{13}$, even in a conservative scenario where a composition dependence of acceleration rates masks a gravitational self energy dependence, $\eta$ is constrained to be less than 0.0008 \cite{[2]}; without such accidental cancelation the $\eta$ constraint improves to 0.0003. The most precise value for the PPN parameter  $\gamma$ is at present given by Bertotti et al \cite{cassini_ber} as: $\gamma -1 = (2.1\pm2.3)\times 10^{-5}$, which was obtained from a solar conjunction experiment with the Cassini spacecraft.   

We shall now discuss motivations for the precision gravity tests that recently became available from both theory and experiment. 

\subsection{ Motivations for Precision Gravity Experiments} 
\label{sec:mot}

Almost ninety years after general relativity was born, Einstein's theory has survived every test. Such a longevity, along with the absence of any adjustable parameters, does not mean that this theory is absolutely correct, but it serves to motivate more accurate tests to determine the level of accuracy at which it is violated. A significant number of these tests  were conducted over the period of the last 35 years. As an upshot of these efforts, most alternative theories have been put aside; only those theories of gravity flexible enough have survived, the accommodation being provided by free parameters and coupling constants of the theory.  

Recently considerable interest has been shown in the physical processes occurring in the strong gravitational field regime. It should be noted that general relativity and some other alternative gravitational theories are in good agreement with the experimental data collected from the relativistic celestial mechanical extremes provided by the relativistic motions in the binary millisecond pulsars.  However, many modern theoretical models, which include general relativity as a standard gravity theory, are faced with the problem of the unavoidable appearance of space-time singularities. It is generally suspected that the classical description, provided by general relativity, breaks down in a domain where the curvature is large, and, hence, a proper understanding of such regions requires new physics. 

The continued inability to merge gravity with quantum mechanics indicate that the pure tensor gravity of general relativity needs modification or augmentation. The tensor-scalar theories of gravity, where the usual general relativity tensor field coexists with one or several long-range scalar fields, are believed to be the most promising extension of the theoretical foundation of modern gravitational theory. The superstring, many-dimensional Kaluza-Klein, and inflationary cosmology theories have revived interest in the so-called `dilaton fields', i.e. neutral scalar fields whose background values determine the strength of the coupling constants in the effective four-dimensional theory. The importance of such theories is that they provide a possible route to the quantization of gravity. Although the scalar fields naturally appear in the theory, their inclusion predicts different relativistic corrections to Newtonian motions in gravitating systems. These deviations from GR lead to a violation of the Equivalence Principle (either weak or strong or both), modification of large-scale gravitational phenomena, and generally lead to space and time variation of physical `constants.' As a result, this progress provides new strong motivation for high precision relativistic gravity tests.

\begin{figure}[t!]
 \begin{center}
\noindent    
\psfig{figure=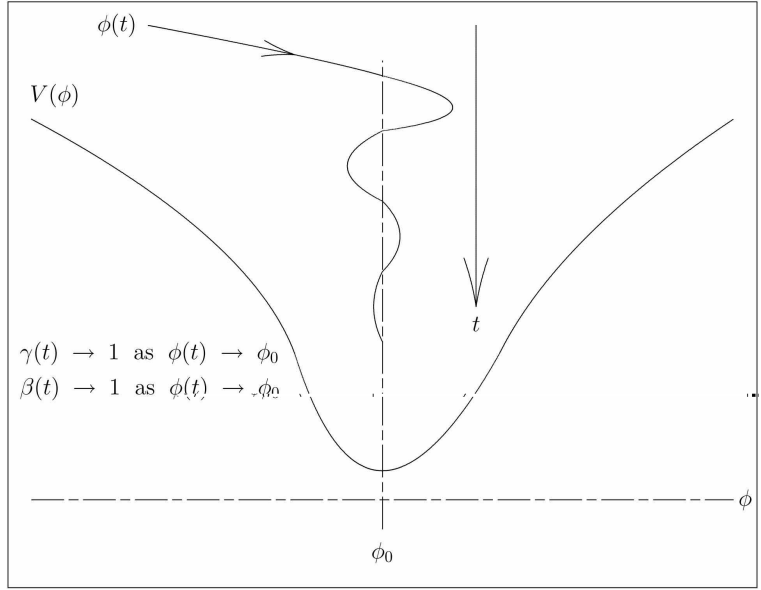,width=100mm}
\end{center}
\vskip -10pt 
  \caption{Typical cosmological dynamics of a background scalar field is shown if that field's coupling function to matter, $V(\phi)$, has an attracting point $\phi_0$. The strength of the scalar interaction's coupling to matter is proportional to the derivative (slope) of the coupling function, so it weakens as the attracting point is approached, and both the Eddington parameters $\gamma$ and $\beta$ (and all higher structure parameters as well)  approach their pure tensor gravity values in this limit.  However, a small residual scalar gravity should remain today because this dynamical process is not complete, and that is what experiment seeks to find.
 \label{fig:attract}}
\end{figure}

The recent theoretical findings suggest that the present agreement between Einstein's theory and experiment might be naturally compatible with the existence of a scalar contribution to gravity. In particular, Damour and Nordtvedt \cite{[12a]} (see also \cite{[10],[15]} for non-metric versions of this mechanism) have recently found that a scalar-tensor theory of gravity may contain a `built-in' cosmological attractor mechanism towards GR.  A possible scenario for cosmological evolution of the scalar field was given in \cite{[12a],[12n]}. Their speculation assumes that the parameter  $\frac{1}{2}(1-\gamma)$  was of order 1 in the early universe, at the time of inflation, and has evolved to be close to, but not exactly equal to, zero at the present time (Figure \ref{fig:attract} illustrates this mechanism in more detail). The expected deviation from zero may be of the order of the inverse of the redshift of the time of inflation, or somewhere between 1 part per $10^5$ and 1 part per $10^7$ depending on the total mass density of the universe:  $1-\gamma \sim 7.3 \times 10^{-7}(H_0/\Omega_0^3)^{1/2}$, where $\Omega_0$ is the ratio of the current density to the closure density and $H_0$ is the Hubble constant in units of 100 km/sec/Mpc. This recent work in scalar-tensor extensions of gravity which are consistent with, indeed often part of, present cosmological models motivates new searches for very small deviations of relativistic gravity in the solar system, at levels of 10$^{-5}$ to 10$^{-7}$ of the post-Newtonian effects.   

The theoretical arguments above have been unexpectedly joined by a number of experimental results that motivate more precise gravitational experiments. In particular, there is now multiple evidence indicating that 70\% of the critical density of the universe is in the form of a `negative-pressure' dark energy component; there is no understanding as to its origin and nature. The fact that the expansion of the universe is currently undergoing a period of acceleration now seems inescapable: it is directly measured from the light-curves of several hundred type Ia supernovae \cite{[2c],[52],[3c]}, the masses of large-scale structures \cite{[29]}, and independently inferred from observations of CMB (Cosmic Microwave Background) by the WMAP satellite \cite{[4c]} and other CMB experiments \cite{[20],[6c],[7c]}. Cosmic speed-up can be accommodated within general relativity by invoking a mysterious cosmic fluid with large negative pressure, dubbed dark energy. The simplest possibility for dark energy is a cosmological constant; unfortunately, the smallest estimates for its value are 55 orders-of-magnitude too large (for reviews see \cite{[7c]} and references therein). 

Most of the theoretical studies operate in the shadow of the cosmological constant problem, the most embarrassing hierarchy problem in physics. This fact has motivated a host of other possibilities, most of which assume $\Lambda=0$, with the dynamical dark energy being associated with a new scalar field. The implication of these observations for cosmological models is that a classically evolving scalar field currently dominates the energy density of the universe. Such models have been shown to share the advantages of  $\Lambda$:  compatibility with the spatial flatness predicted inflation; a universe older than the standard Einstein-de Sitter model; and, combined with cold dark matter, predictions for large-scale structure formation in good agreement with data from galaxy surveys.  Combined with the fact that scalar field models imprint distinctive signature on CMB anisotropy, they remain currently viable and should be testable in the near future. 
On the other hand, none of these suggestions is very compelling and most have serious drawbacks. Given the challenge of this problem, a number of authors considered the possibility that cosmic acceleration is not due to some kind of stuff, but rather arises from new gravitational physics (see discussion in \cite{[carroll]}). In particular, extensions to general relativity in a low curvature regime were shown to predict an experimentally consistent universe evolution  without the need for dark energy. These dynamical models are expected to produce measurable contribution to the parameter $\gamma$  in experiments conducted in the solar system also at the level of $1-\gamma \sim 10^{-7}-10^{-9}$, thus further motivating the relativistic gravity research. Therefore, the PPN parameter $\gamma$ may be the only key parameter that holds the answer to most of the questions discussed. 

This completely unexpected discovery demonstrates the importance of testing the important ideas about the nature of gravity. We are presently in the `discovery' phase of this new physics, and while there are many theoretical conjectures as to the origin of a non-zero $\Lambda$, it is essential that we exploit every available opportunity to elucidate the physics that is at the root of the observed phenomena. There is also experimental evidence for time-variability in the fine structure constant, $\alpha$, at the level of $\dot\alpha/(\alpha H_0) \sim 10^{-5}$  \cite{[34]}. This is very similar to time variation in the gravitational constant, which at the post-Newtonian level is expressed as $\dot G/(G H_0)\approx \eta = 4\beta-\gamma-3$, thus providing a tantalizing motivation for further tests of the SEP parameter $\eta$. A similar conclusion resulted from the recent analysis performed in \cite{[1],[7],[9]}. These new findings necessitate the measurements of $\gamma$  and $\beta$ in the range from $10^{-6}$ to $10^{-8}$ to test the corresponding gravitational scenario, thus requiring new gravitational physics missions. 

In summary, there are a number of theoretical reasons to question the validity of GR. Despite the success of modern gauge field theories in describing the electromagnetic, weak, and strong interactions, it is still not understood how gravity should be described at the quantum level. In theories that attempt to include gravity, new long-range forces can arise in addition to the Newtonian inverse-square law. Even at the purely classical level, and assuming the validity of the Equivalence Principle, Einstein's theory does not provide the most general way to generate the space-time metric. Regardless of whether the cosmological constant should be included, there are also important reasons to consider additional fields, especially scalar fields.  Also, the recent accuracy improvement in tests of gravity in the solar system is not sufficient to lead to groundbreaking tests of fundamental physical laws addressed above. This is especially true if the cosmological attractor discovered in \cite{[12a],[12n]} is more robust, time variation in the fine structure constant would be confirmed in other experiments and various GR extensions would demonstrate feasibility of these methods for cosmology and relativistic gravity. 

The new LLR capabilities and the proposed LATOR mission are poised to directly address the challenges discussed above; we shall now discuss these experiments in more details.

\section{Lunar Laser Ranging: a Unique Laboratory in Space}
  
\subsection{LLR History and Scientific Background}

LLR has a distinguished history \cite{[15]} dating back to the placement of retroreflector arrays on the lunar surface by the Apollo 11 astronauts. Additional reflectors were left by the Apollo 14 and Apollo 15 astronauts, and two French-built reflector arrays were placed on the Moon by the Soviet Luna 17 and Luna 21 missions. Figure 2 shows the weighted RMS residual of laser ranges to these reflector arrays for each year.  Early accuracies using the McDonald Observatory's 2.7 m telescope hovered around 25 cm.  Equipment improvements decreased the ranging uncertainty to $\sim$15 cm later in the 1970s.  In 1985 the 2.7 m ranging system was replaced with the McDonald Laser Ranging System (MLRS).  In the 1980s ranges were also received from Haleakala Observatory on the island of Maui in the Hawaiian chain and the Observatoire de la Cote d'Azur (OCA) in France.  Haleakala ceased operations in 1990.  A sequence of technical improvements decreased the range uncertainty to the current $\sim$ 2 cm level.  The 2.7 m telescope had a greater light gathering capability than the newer smaller aperture systems, but the newer systems fired more frequently and had a much improved range accuracy.  The new systems cannot distinguish returning photons against the bright background near full Moon, which the 2.7 m telescope could do,  though there are some modern eclipse observations at full moon.  

LLR accurately measures the time of flight for a laser pulse fired from an observatory on the Earth, bounced off of a corner cube retroreflector on the Moon, and returned to the observatory.  For a general review of LLR see Dickey et al. \cite{[15]}.  A comprehensive paper on tests of gravitational physics is Williams et al. \cite{[62]}. A recent test of the Equivalence Principle is in Anderson and Williams \cite{[2]} and other gravitational physics tests are in Williams et al. \cite{jim01}.  An overview of the LLR gravitational physics tests is given by Nordtvedt \cite{ken30}.  Reviews of various tests of relativity, including the contribution by LLR, are given in Will \cite{[58]}. 

The LLR measurements of the past have contributed to a wide range of scientific investigations \cite{[62],[2],[7]}, and are today solely responsible for the production of the lunar ephemeris. On the fundamental scientific front, LLR provides the only means for testing the SEP---the statement that \emph{all} forms of mass and energy contribute equivalent quantities of inertial and gravitational mass. In addition, LLR is capable of measuring the time variation of Newton's gravitational constant, {\it G}, providing the strongest limit available for the variability of this `constant'. LLR can also precisely measure the de Sitter precession---effectively a spin-orbit coupling affecting the lunar orbit in the frame co-moving with the Earth-Moon system's motion around the Sun. Finally, current LLR results are consistent with the existence of  gravitomagnetism within 0.1\% of the predicted level \cite{[12n],ken30}, thus making the lunar orbit a unique laboratory for gravitational physics where each term in the relativistic equations of motion has been verified to a very high accuracy. Besides the fundamental physics capabilities of LLR, the interior, tidal response, and physical librations (rocking) of the Moon are all probed by LLR, making it a valuable tool for physical selenography \cite{jj02}.

\begin{figure}[!t]
 \begin{center}
\noindent    
\psfig{figure=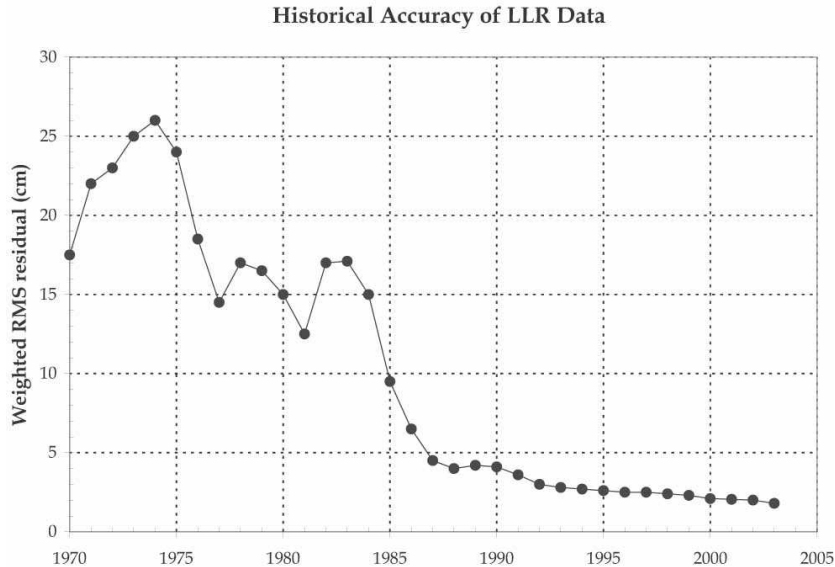,width=110mm}
\end{center}
\vskip -10pt 
  \caption{Historical accuracy of LLR data from 1970 to 2003.  
 \label{accuracy}}
\end{figure} 


The APOLLO lunar laser-ranging project will yield a one order-of-magnitude improvement in the precision of three important tests of the basic properties of the gravitational interaction.  Below we shall discuss some expected results and their significance for fundamental and gravitational physics.

\subsection{Equivalence Principle Tests}

The Equivalence Principle, the exact correspondence of gravitational and inertial masses, is a central assumption of general relativity and a unique feature of gravitation. It is the equivalence principle that leads to identical accelerations of compositionally different objects in the same gravitational field, and also allows gravity to be viewed as a geometrical property of spacetime---leading to the general relativistic interpretation of gravitation. EP tests can therefore be viewed in two contexts: tests of the foundations of the Standard Model of Gravity (i.e. general relativity), or as searches for new physics because, as emphasized in \cite{[12a],[10],[15],[12n]}, almost all extensions to the Standard Model of particle physics generically predict new forces that would show up as apparent violations of the EP. Easily the most precise tests of the EP are made by simply comparing the free fall accelerations, $a_1$ and $a_2$, of different test bodies, with
\begin{equation}
\frac{\Delta a}{a}\equiv \frac{2(a_1-a_2)}{(a_1+a_2)}=\left(\frac{M_G}{M_I}\right)_1-\left(\frac{M_G}{M_I}\right)_2
\end{equation}
where $M_G$  and $M_I$  represent gravitational and inertial masses of each body. The sensitivity of the EP test is determined by the precision of the differential acceleration measurement divided by the degree to which the test bodies differ (e.g. composition).

\subsubsection{The Weak Equivalence Principle}
The weak form the EP (the WEP) states that the gravitational properties of strong and electro-weak interactions obey the EP. In this case the relevant test-body differences are their fractional nuclear-binding differences, their neutron-to-proton ratios, their atomic charges, etc. General relativity, as well as other metric theories of gravity, predict that the WEP is exact. However, extensions of the Standard Model of Particle Physics that contain new macroscopic-range quantum fields predict quantum exchange forces that will generically violate the WEP because they couple to generalized `charges' rather than to mass/energy as does gravity \cite{[15]}. WEP tests can be conducted with laboratory or astronomical bodies, because the relevant differences are in the test-body compositions.

\subsubsection{The Strong Equivalence Principle}
The strong form of the EP extends the principle to cover the gravitational properties of gravitational energy itself. In other words it is an assumption about the way that gravity begets gravity, i.e. about the non-linear property of gravitation. Although general relativity assumes that the SEP is exact, alternate metric theories of gravity such as those involving scalar fields, and other extensions of gravity theory, typically violate the SEP \cite{[39],[12n]}. For the SEP case, the relevant test body differences are the fractional contributions to their masses by gravitational self-energy. Because of the extreme weakness of gravity, SEP test bodies that differ significantly must have astronomical sizes. Currently the Earth-Moon-Sun system provides the best arena for testing the SEP.

To facilitate investigation of a possible violation of the SEP, the ratio between gravitational and inertial masses, $M_G/M_I$  is expressed in the form
\begin{equation}
\frac{M_G}{M_I}=1+\eta\frac{U}{Mc^2}
\end{equation}	  	

\noindent where $U$ is the gravitational self-energy of the body $(U < 0)$, $Mc^2$ is its total mass-energy, and $\eta$  is a dimensionless constant. $U/Mc^2$  is proportional to $M$, so testing the SEP requires bodies the size of the Moon and planets. For the Earth-Moon system,
\begin{equation}
\frac{U_e}{M_ec^2}-\frac{U_m}{M_mc^2}=-4.45\times 10^{-10}
\end{equation}	
where the subscripts $e$ and $m$ denote the Earth and Moon, respectively. Therefore, a violation of the SEP would produce an Earth-Moon differential acceleration of $\Delta a/a =  -4.45\times10^{-10}\eta $.

In general, $\eta$  is a linear function of seven of the ten Parameterized Post-Newtonian (PPN) parameters, but considering only $\beta$  and  $\gamma$
\begin{equation}
\eta=4\beta-\gamma-3
\label{3}
\end{equation}
In general relativity $\eta  = 0$. A unit value for $\eta$  would produce a displacement of the lunar orbit about the Earth \cite{ken_dm2,damour_vokr}, causing a 13 meter monthly range modulation.

\subsection{LLR Tests of the Equivalence Principle}
\label{sec:EP}

In essence, LLR tests of the EP compare the free-fall accelerations of the Earth and Moon toward the Sun. Lunar laser-ranging measures the time-of-flight of a laser pulse fired from an observatory on the Earth, bounced off of a retroreflector on the Moon, and returned to the observatory \cite{[15llr],[9]}. If the Equivalence Principle is violated, the lunar orbit will be displaced along the Earth-Sun line, producing a range signature having a 29.53 day synodic period (different from the lunar orbit period of 27 days). Since the first LLR tests of the EP were published in 1976 \cite{[54],jj02,[17]}, the precision of the test has increased by two orders-of-magnitude \cite{[62],[2],jj02,[20]}. (Reviews of contributions to gravitational physics by LLR are given by Nordtvedt \cite{ken30} and Will \cite{[58]}.)

From the viewpoint of the EP, the Earth and Moon `test bodies' differ in two significant ways: in composition (the Earth has a massive Fe/Ni core while the Moon has a much smaller core) and in their gravitational self-energies (the Earth is much more massive than the Moon). Therefore, LLR tests the total Equivalence Principle---composition plus self-energy---for the Earth and Moon in the gravitational field of the Sun. Two recent results yield  $\Delta a/a$ values of $(-1 \pm 2)\times 10^{-13}$  \cite{jim01} and $(-0.7 \pm 1.5)\times 10^{-13}$ \cite{ken30}. The latter corresponds to a $2\pm4$ mm amplitude in range.

The LLR result is a null test so it can be argued that it is unlikely that there would be two compensating violations of the Equivalence Principle---composition and self-energy---that essentially cancel. However, because of the fundamental importance of a good SEP test, laboratory tests of the WEP are used to separate with certainty any composition-dependent and self-energy effects. Recent WEP tests performed at the University of Washington (UW) using laboratory test bodies whose compositions are close to those of the actual Earth and Moon set upper limits on any composition-dependent Earth-Moon differential acceleration \cite{[1],[7]}. The random and systematic  $\Delta a/a$ uncertainties of \cite{[1]} are $1.4\pm10^{- 13}$  and  $0.2\times10^{-13}$, respectively. Anderson and Williams \cite{[2]} used the earlier of these WEP results \cite{[7]} to limit the SEP parameter  $\eta = 0.0002\pm0.0008$. If one adopts the more recent WEP test by the UW E\"ot-Wash group \cite{[1]}, one gets an $\eta$ uncertainty of 0.0005. Note that the current intrinsic LLR accuracy, if the WEP were known perfectly, is 0.0003. Therefore, with its 1 mm range accuracy, APOLLO has the capability of determining  $\eta$ to a precision of approximately $3\times 10^{-5}$.

\subsection{LLR Tests of Other Gravitational Physics Parameters}
In addition to the SEP constraint based on Eq.(\ref{3}), the PPN parameters  $\gamma$ and $\beta$  affect the orbits of relativistic point masses, and $\gamma$  also influences time delay \cite{[62]}. LLR tests this $\beta$  and  $\gamma$ dependence, as well as geodetic precession, and $\dot G/G$.  The possibility of a time variation of the constant of gravitation, {\it G}, was first considered by Dirac in 1938 on the basis of his large number hypothesis, and later developed by Brans and Dicke  in their theory of gravitation (for more details consult \cite{[59]}). Variation could be related to the expansion of the Universe, in which case $\dot G/G=\sigma H_0$, where $H_0$ is the Hubble constant, and $\sigma$ is a dimensionless parameter whose value depends on both the gravitational constant and the cosmological model considered. Revival of interest in the Brans-Dicke-like theories, with a variable {\it G}, was partially motivated by the appearance of superstring theories where {\it G} is considered to be a dynamical quantity \cite{Marciano1984}). A scale-dependent gravitational constant could mimic the presence of dark matter \cite{Goldman1992} and could enter discrepancies between the determinations of $H_0$ at different scales \cite{Bertolami}.
Williams et al. \cite{jim01} give uncertainties of 0.004 for  $\beta$ and $\gamma$  deduced from sensitivity apart from the SEP, and $1.1\times 10^{- 12}$  yr$^{-1}$ for $\dot G/G$ test. 

The SEP relates to the non-linearity of gravity (how gravity affects itself), with the PPN parameter  $\beta$ representing the degree of non-linearity. Thus LLR provides the best way to measure  $\beta$, as suggested by the strong dependence of $\eta$ on  $\beta$  in Eq. (\ref{3}). The parameter $\gamma$  has been measured independently via time-delay and gravitational ray-bending techniques. The published Viking \cite{[47]} and Very Long-Baseline Interferometry (VLBI) \cite{[48]} uncertainties for $\gamma$  are 0.002, 0.002, and 0.0022, respectively. Combining the above limits on  $\eta$  from LLR and laboratory WEP tests with the Viking and VLBI results for  $\gamma$  gives  $|\beta-1|<0.0005$, the limit given by \cite{[2]}. The uncertainty in $\beta$  determined in this way is dominated by the uncertainty in $\gamma$. Fortunately, a much more accurate result for $\gamma$ was recently reported by the Cassini experiment \cite{cassini_ber}; this leads to a significant improvement in the parameter $\beta$ determination. 

In our recent LLR analysis with data to May 2003, the Equivalence Principle was tested at the level of $M_G/M_I =(0.5\pm1.4)\times10^{-13}$, including correction for solar radiation pressure. This result corresponds to the SEP test at the level of $\Delta a/a =(-1.5\pm2.0)\times10^{-13}$ (with a WEP result from \cite{[1]}) and $\eta=(3.4\pm4.5)\times 10^{-4}$ for the SEP violation parameter. Using the Cassini result for  $\gamma$ from \cite{cassini_ber}, the PPN parameter $\beta$ was measured at the level of $\beta=1+(0.9\pm1.1)\times 10^{-4}$. The geodetic precession was tested at the level of $K_{gp}=-0.0035\pm0.0066$ and the search for variation in gravitational constant resulted in $\dot G/G=(0.46\pm1.0)\times10^{-12}$ yr$^{-1}$.

Orbital precession depends on $\beta$  and $\gamma$, so their sensitivity depends on the time span of the data. The uncertainty for $\dot G/G$ is improving rapidly because its sensitivity depends on the square of the time span. So 1 mm quality data would improve the $G$ rate uncertainty by an order-of-magnitude in $\sim$ 5 yr while $\gamma$  and geodetic precession would depend on orbital precession time scales: 6.0 yr for argument of perigee, 8.85 yr for longitude of perigee, and 18.6 yr for node.

LLR also has the potential to determine the solar $J_2$ \cite{jim01}, PPN  $\alpha_1$ \cite{damour_vokr,muller}, hunt for influences of dark matter \cite{ken_dm2,ken_dm1}, and to test the inverse square law at the scale of $ae \sim$  20,000 km.  A long-range Yukawa interaction has been tested by M\"uller et al. \cite{MG7}.  

\subsection{APOLLO Contribution to the Tests of Gravity}

The Apache Point Observatory Lunar Laser-ranging Operation is a new LLR effort designed to achieve millimeter range precision and corresponding order-of-magnitude gains in measurements of fundamental physics parameters. The APOLLO project design and leadership responsibilities are shared between the University of California at San Diego and the University of Washington. In addition to the modeling aspects related to this new LLR facility, a brief description of APOLLO and associated expectations is provided here for reference. A more complete description can be found in \cite{[35],[35a]}.

The overwhelming advantage APOLLO has over current LLR operations is a 3.5 m astronomical quality telescope at a good site. The site in the Sacramento Mountains of southern New Mexico offers high altitude (2780 m) and very good atmospheric `seeing' and image quality, with a median image resolution of 1.1 arcseconds. Both the image sharpness and large aperture enable the APOLLO instrument to deliver more photons onto the lunar retroreflector and receive more of the photons returning from the reflectors, respectively. Compared to current operations that receive, on average, fewer than 0.01 photons per pulse, APOLLO should be well into the multi-photon regime, with perhaps 5-10 return photons per pulse. With this signal rate, APOLLO will be efficient at finding and tracking the lunar return, yielding hundreds of times more photons in an observation than current operations deliver. In addition to the significant reduction in statistical error ($\sim\sqrt{N}$ reduction), the high signal rate will allow assessment and elimination of systematic errors in a way not currently possible.

The new LLR capabilities introduced by APOLLO offer a unique opportunity to improve the accuracy of a number of fundamental physics tests. Some of them would have a profound effect on our understanding of the evolution of our universe.  If {\it G} changes at a rate comparable to the reported change in the fine structure constant ($\dot\alpha/\alpha \sim 10^{-15}$ yr$^{- 1}$) \cite{[29]}, $\eta$  would be approximately $10^{- 5}$. Thus, an order-of-magnitude LLR range improvement would give an   uncertainty within reach of the predictions by Damour and Nordtvedt ($\sim 10^{-7} < \eta  < 10^{- 4}$ \cite{[12a]}), and comparable to the value implied by $\dot\alpha$ ($\dot\alpha/\alpha \sim 10^{-15}$ yr$^{- 1}$ \cite{[34]}).

The APOLLO project will push LLR into the regime of millimetric range precision which translates to an order-of-magnitude improvement in the determination of fundamental physics parameters. For the Earth and Moon orbiting the Sun, the scale of relativistic effects is set by the ratio $(GM / r c^2)\sim v^2 /c^2 \sim 10^{-8}$.  Relativistic effects are small compared to Newtonian effects.  The Apache Point 1 mm range accuracy corresponds to $3\times 10^{-12}$ of the Earth-Moon distance.  The resulting LLR tests of gravitational physics would improve by an order-of-magnitude: the Equivalence Principle would give uncertainties approaching $10^{-14}$, tests of general relativity effects would be $<0.1$\%, and estimates of the relative change in the gravitational constant would be 0.1\% of the inverse age of the universe. This last number is impressive considering that the expansion rate of the universe is approximately one part in 10$^{10}$ per year.

\section{New Test of Relativity: The LATOR Mission} 

The technology has advanced to the point that one can consider carrying out direct tests in a weak field to second order in the field strength parameter $\propto GM/rc^2$. Although any measured anomalies in first or second order metric gravity potentials will not determine strong field gravity, they would signal that modifications in the strong field domain exist.  The converse is perhaps more interesting:  if to high precision no anomalies are found in the lowest order metric potentials, and this is reinforced by finding no anomalies at the next order, then it follows that any anomalies in the strong gravity environment are correspondingly quenched. This topic will be the main science goal of the LATOR mission.

\subsection{Overview of LATOR} 
\label{sec:lator_description}

\begin{figure*}[t!]
 \begin{center}
\noindent    
\psfig{figure=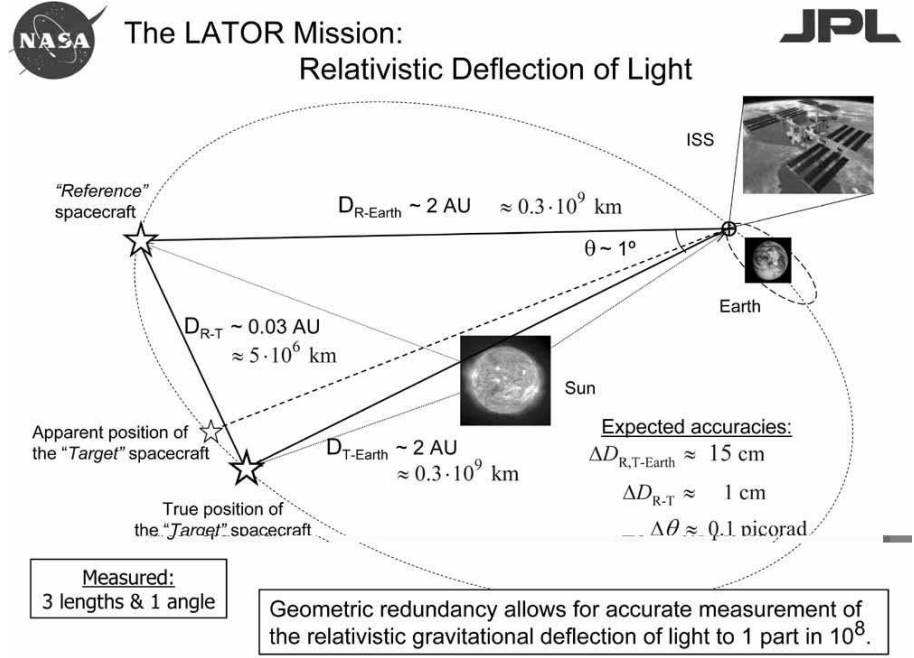,width=120mm}
\end{center}
\vskip -10pt 
  \caption{Geometry of the LATOR experiment to measure deviations from the Euclidean geometry in the solar gravity field. 
 \label{fig:lator}}
\end{figure*}

The LATOR experiment would use laser interferometry between two micro-spacecraft (placed in heliocentric orbits, at distances $\sim$ 1 AU from the Sun), whose lines of sight pass close by the Sun, to accurately measure deflection of light in the solar gravity. Another component of the experimental design is a long-baseline ($\sim100$ m) multi-channel stellar optical interferometer placed on the International Space Station (ISS). Figure \ref{fig:lator} shows the general concept for the LATOR missions including the mission-related geometry, experiment details  and required accuracies. 

The LATOR mission consists of two low cost micro-spacecraft (the goal is to launch both spacecraft on a single Delta II launch vehicle). with three interferometric links between the craft and a beacon station on the ISS.  One of the longest arms of the triangle ($\sim$ 2 AU) passes near the Sun. The two spacecraft are in heliocentric orbits and use lasers to measure the distance between themselves and a beacon station on the ISS. The laser light passes close to the Sun, which causes the light path to be both bent and lengthened. One spacecraft is at the limb of the Sun, the other one is $\sim 1^\circ$ away, as seen from the ISS. Each spacecraft uses laser ranging to measure the distance changes to the other spacecraft. The spatial interferometer is for measuring the angles between the two spacecraft and for orbit determination purposes.

As evident from Figure \ref{fig:lator}, the key element of the LATOR experiment is a redundant geometry optical truss to measure the departure from Euclidean geometry caused by gravity.  The triangle in figure has three independent arms the lengths of which are monitored with laser metrology. From three measurements one can calculate the Euclidean value for any angle in this triangle.  In Euclidean geometry these measurements of of the three lengths of the triangle should agree with the angle measured by the interfereometer to high accuracy.  This geometric redundancy enables LATOR to measure the departure from Euclidean geometry caused by the solar gravity field to a very high accuracy. The difference in the measured angle and its Euclidean value is the non-Euclidean signal. To avoid having to make absolute measurements, the spacecraft are placed in an orbit where their impact parameters, the distance between the beam and the center of the Sun, vary significantly from $10 R_\odot$ to $1 R_\odot$ over a period of $\sim$ 20 days.

The shortening of the interferometric baseline is achieved solely by going into space to avoid the atmospheric turbulence and Earth's seismic vibrations. On the space station, all vibrations can be made common mode for both ends of the interferometer by coupling them by an external laser truss. This relaxes the constraint on the separation between the spacecraft, allowing it to be as large as a few degrees as seen from the ISS. Additionally, the orbital motion of the ISS provides variability in the interferometer's baseline projection as needed to resolve the fringe ambiguity of the stable laser light detection by an interferometer.

\subsection{The Expected Results from LATOR} 

The first order effect of light deflection in the solar gravity caused by the solar mass monopole is 1.75 arcseconds (see Table \ref{tab:eff} for more details), which corresponds to a delay of $\sim$0.85 mm on a 100 m baseline. We currently are able to measure with laser interferometry distances with an accuracy (not just precision but accuracy) of $<$ 1 picometer. In principle, the 0.85 mm gravitational delay can be measured with $10^{-9}$ accuracy versus $10^{-4}$ available with current techniques. However, we use a conservative estimate for the delay of 10 pm which would produce the measurement of $\gamma$ to accuracy of 1 part in $10^{-8}$ (i.e improving the accuracy in determining this parameter by a factor of 30,000) rather than 1 part in $10^{-9}$. Note that the Eddington parameter $\gamma$, whose value in general relativity is unity, is perhaps the most fundamental PPN parameter, in that $(1-\gamma)$ is a measure, for example, of the fractional strength of the scalar gravity interaction in scalar-tensor theories of gravity.  Within perturbation theory for such theories, all other PPN parameters to all relativistic orders collapse to their general relativistic values in proportion to $(1-\gamma)$. Therefore, measurement of the first order light deflection effect at the level of accuracy comparable with the second-order contribution would provide the crucial information separating alternative scalar-tensor theories of gravity from general relativity \cite{[12a],[12n]}. 

\begin{table*}[!t!]
\caption{Comparable sizes of various light deflection effects in the solar gravity field.\label{tab:eff}}
\begin{center}
\begin{tabular}{|c|c|c|c|} \hline &&&\\[-8pt]
     Effect  & 
     Analytical Form      & 
     Value ($\mu$as)  &
     Value (pm) \\[3pt]
\hline \hline
&&&\\[-9pt]
   First  Order  &
   $2(1+\gamma )\frac{GM}{c^2R}$ & 
   $1.75\times 10^6$   & 
   $8.487\times10^{8}$      \\[4pt] 
&&&\\[-12pt] \hline 
&&&\\[-9pt]
   Second Order   &
   $[(2(1+  \gamma)-\beta+\frac{3}{4} \delta)\pi- 
2(1+\gamma)^2](\frac{GM}{c^2R})^2 $
   & 3.5   
   & 1702      \\[4pt] 
&&&\\[-12pt] \hline 
&&&\\[-8pt]
   Frame-Dragging   &
   $\pm 2(1+\gamma)\frac{GJ}{c^3R^2}$ & 
   $\pm0.7$   &     $\pm339$ \\[4pt] 
&&&\\[-12pt] \hline 
&&&\\[-8pt]
   Solar Quadrupole  &
   $2(1+\gamma )J_2\frac{GM}{c^2R}$ & 
   0.2   &   97  \\[4pt] 
\hline 
\end{tabular}
\end{center}   

\end{table*}


Where the light deflection by solar gravity is concerned, the magnitude of the first order effect as predicted by GR for the light ray just grazing the limb of the Sun is $\sim1.75$ arcsecond (consult Table \ref{tab:eff}). The effect varies inversely with the impact parameter. The second order term is almost six orders-of-magnitude smaller resulting in  $\sim 3.5$ microarcseconds ($\mu$as) light deflection effect, and it falls off inversely as the square of the light ray's impact parameter \cite{[22s],[26s],[49s],[40s]}.
The relativistic frame-dragging term\footnote{Gravitomagnetic frame dragging is the effect in which both the orientation and trajectory of objects in orbit around a body are altered by the gravity of the body's rotation.  It was studied by Lense and Thirring in 1918.} is $\pm 0.7 ~\mu$as, and the contribution of the solar quadrupole moment, $J_2$, is sized as 0.2 $\mu$as (using the value of the solar quadrupole moment $J_2\simeq10^{-7}$ ). The small magnitudes of the effects emphasize the fact that, among the four forces of nature, gravitation is the weakest interaction; it acts at very long distances and controls the large-scale structure of the universe, thus making the precision tests of gravity a very challenging task. 

 The second order light deflection is approximately 1700 pm and with 10 pm accuracy it could be measured with accuracy of $\sim1\times 10^{-3}$, including first ever measurement of the PPN parameter $\delta$.  The frame dragging effect would be measured with $\sim 1\times10^{-2}$ accuracy and the solar quadrupole moment (using the theoretical value of the solar quadrupole moment $J_2\simeq10^{-7}$) can be modestly measured to 1 part in 20, all with respectable signal to noise ratios.

The laser interferometers use $\sim$2W lasers and $\sim$20 cm optics for transmitting the light between spacecraft. Solid state lasers with single frequency operation are readily available and are relatively inexpensive.   For SNR purposes we assume the lasers are ideal monochromatic sources. For simplicity we assume the lengths being measured are 2AU = $3\times 10^8$ km. The beam spread is 1 $\mu$m/20 cm = 5 $\mu$rad (1 arcsecond). The beam at the receiver is $\sim$1,500 km in diameter, a 20 cm receiver will detect $1.71 \times 10^2$ photons/sec assuming 50\% q.e. detectors. 5 picometer (pm) resolution for a measurement of $\gamma$ to $\sim10^{-8}$ is possible with approximately 10 seconds of integration.

As a result, the LATOR experiment will be capable of measuring the angle between the two spacecraft to $\sim 0.01 ~\mu$as, which allows light deflection due to gravitational effects to be measured to one part in $10^8$. Measurements with this accuracy will lead to a better understanding of gravitational and relativistic physics. In particular, with LATOR, measurements of the first order gravitational deflection will be improved by a factor of 30,000. LATOR will also be capable of distinguishing between first order ($\propto  GM/c^2R$) and second order ($\propto (GM/c^2R)^2$) effects. All effects, including the first and second order deflections, as well as the frame dragging component of gravitational deflection and the quadrupole deflection, will be measured astrometrically.  

The LATOR experiment has a number of advantages over techniques which use radio waves to measure gravitational light deflection. Advances in optical communications technology, allow low bandwidth telecommunications with the LATOR spacecraft without having to deploy high gain radio antennae needed to communicate through the solar corona. The use of monochromatic light enables the observation of the spacecraft almost at the limb of the Sun, as seen from the ISS. The use of narrowband filters, coronagraph optics and heterodyne detection will suppress background light to a level where the solar background is no longer the dominant noise source. In addition, the short wavelength allows much more efficient links with smaller apertures, thereby eliminating the need for a deployable antenna. Finally, the use of the ISS will allow conducting the test above the Earth's atmosphere---the major source of astrometric noise for any ground based interferometer. These facts justify LATOR as a space mission.

The LATOR experiment technologically is a very sound concept; all technologies that are needed for its success have been already demonstrated as a part of the JPL's Space Interferometry Mission development.  The concept arose from several developments at NASA and JPL that initially enabled optical astrometry and metrology, and also led to developing expertise needed for the precision gravity experiments. Technology that has become available in the last several years, such as low cost microspacecraft, medium power highly efficient solid state lasers for space applications, and the development of long range interferometric techniques, make the LATOR mission feasible. The LATOR experiment does not need a drag-free system, but uses a geometric redundant optical truss to achieve a very precise determination of the interplanetary distances between the two micro-spacecraft and a beacon station on the ISS. The interest of the approach is to take advantage of the existing space-qualified optical technologies leading to an outstanding performance in a reasonable mission development time.  The  availability of the space station makes this mission concept realizable in the very near future; the current mission concept calls for a launch as early as in 2009 with a cost of a NASA MIDEX mission.   

\section{Conclusions} 
\label{sec:conc}
 
LLR provides the most precise way to test the EP for gravity itself, the best way to test for both non-gravitational long-range fields of dark matter as well as for time variation of Newton's constant. With technology improvements and substantial access to a large-aperture, high-quality telescope, the APOLLO project will take full advantage of the lunar retro-reflectors and will exploit the opportunity provided by the unique Earth-Moon `laboratory' for fundamental gravitational physics. The expected improvement in the accuracy of LLR tests of gravitational physics expected with the new APOLLO instrument will bring significant new insights to our understanding of the fundamental physics laws that govern the evolution of our universe. The scientific results are very significant which justifies the more than 35 years of history of LLR research and technology development.  

The LATOR mission aims to carry out a test of the curvature of the solar system's gravity  field with an accuracy better than 1 part in 10$^{8}$. In spite of the previous space missions exploiting radio waves for tracking the spacecraft, this mission manifests an actual breakthrough in the relativistic gravity experiments as it allows one to take full advantage of the optical techniques that have recently become available.  LATOR will lead to very robust advances in the tests of fundamental physics: this mission could discover a violation or extension of general relativity, or reveal the presence of an additional long range interaction in the physical law.  There are no analogs to the LATOR experiment; it is unique and is a natural culmination of solar system gravity experiments. 

\subsubsection*{Acknowledgments~~} 
D.H. Boggs participated in the LLR solutions. The work described here was carried out at the Jet Propulsion Laboratory, California Institute of Technology, under a contract with the National Aeronautics and Space Administration.



\end{document}